\documentclass[12pt]{article}

\usepackage{epsfig}
\usepackage{graphicx}
\usepackage[mathscr]{eucal}

\begin{document}

\title{Phase transition of finite size quark droplets with isospin
chemical potential in the Nanbu--Jona-Lasinio model}

\author{{Guo-yun Shao$^{1,2}$, Lei Chang$^{1}$, Yu-xin Liu$^{1,3,4,5}$,
and Xue-lei Wang$^{2}$ }\\[3mm]
\normalsize{$^1$ Department of Physics, Peking University, Beijing
100871, China}\\[-1mm]
\normalsize{$^{2}$ College of Physics, Henan Normal University,
Xinxiang 453002, China} \\
\normalsize {$^{3}$ Key Laboratory of Heavy Ion Physics, Ministry
of Education, Beijing 100871, China } \\
\normalsize{$^4$ Institute of Theoretical Physics, Academia
Sinica, Beijing 100080, China} \\
\normalsize{$^5$ Center of Theoretical Nuclear Physics, National
Laboratory of}\\ \normalsize{ Heavy Ion Accelerator, Lanzhou
730000, China}  }

%
%\author{Guo-yun Shao}
%\address{College of Physics, Henan Normal University,
%Xinxiang 453002, China} \affiliation{Department of Physics, Peking
%University, Beijing 100871, China}
%
%\author{Lei Chang}
%\address{Department of Physics, Peking University,
%Beijing 100871, China}
%
%\author{Yu-xin Liu}
%\email[corresponding author]{} \affiliation{Department of Physics,
%Peking University, Beijing 100871, China} \affiliation{The Key
%Laboratory of Heavy Ion Physics, Ministry of Education,Beijing
%100871, China }  \affiliation{Center of Theoretical Nuclear
%Physics, National Laboratory of Heavy Ion Accelerator, Lanzhou
%730000, China}
%
%\author{Xue-lei Wang}
%\affiliation{College of Physics, Henan Normal University,
%Xinxiang 453002, China}

\date{\today}

\maketitle

\begin{abstract}
Making use of the NJL model and the multiple reflection expansion
approximation, we study the phase transition of the finite size
droplet with {\it u} and {\it d} quarks. We find that the dynamical
masses of {\it u}, {\it d} quarks are different, and the chiral
symmetry can be restored at different critical radii for {\it u},
{\it d} quark. It provides a clue to understand the effective
nucleon mass splitting in nuclear matter. Meanwhile, it shows that
the maximal isospin chemical potential at zero temperature is much
smaller than the mass of pion in free space.
\end{abstract}

\bigskip

%\pacs{25.75.Nq, 11.10.Wx, 12.38.-t }

{\bf PACS numbers:}  25.75.Nq, 11.10.Wx, 12.38.-t

%\maketitle

\newpage

\parindent=20pt

The stability of quark matter has been investigated recently with
various approaches, such as lattice QCD \cite{KS02,Nishida04}, MIT
bag model \cite{CJT74,Witten84,FJ84,BJ87,Madsen934}, NJL model
\cite{NJL61,VW91,KH03,YHT05,Buballa05}. The strange quark matter
containing {\it u}, {\it d} and {\it s} quarks are considered more
stable than those only containing {\it u}, {\it d} quarks
\cite{Witten84,FJ84}. In the MIT bag model, hadrons consist of free
(or only weakly interacting) quarks which are confined to finite
region of space (in the ``bag''). The confinement is not a dynamical
result of the theory, but an input parameter with appropriate
boundary conditions in the model. However, an important feature of
MIT bag model is the absence of the dynamical chiral symmetry
breaking. When the effect of the dynamical chiral symmetry breaking
is taken into account, the strange quark matter is not absolutely
stable \cite{VW91,BO99,KHK94}. Meanwhile, the transition from {\it
u}, {\it d} quark to {\it u}, {\it d}, {\it s} quark by weak process
is not favored, because when the chiral symmetry of {\it u}, {\it d}
quark is restored, the symmetry of {\it s} quark is still broken
\cite{YHT05}. It is then necessary to study the property of the
system consisting of $u$ and $d$ quarks.

So far all the foregoing analysis are either restricted to the quark
matter that {\it u}, {\it d} quarks have the same chemical potential
and the isospin is symmetric in the Lagrangian, or that of infinite
volume. This means that, for the infinite quark matter, all the
quantities related to {\it u}, {\it d} quarks in the matter are the
same, in particular the dynamical quark mass and the quark
condensates are equal for both flavors. However, there are many
situations in the real world that the {\it u}, {\it d} quark numbers
are not equal. According to the principle of statistical physics,
the chemical potential should be different for {\it u}, {\it d}
quark, respectively. For example, the neutron stars must be
electrically neutral to a very high degree (even though there are
electrons in neutron stars, the ratio is very small)
\cite{Buballa05}. Therefore if the neutron star consists of
deconfined {\it u}, {\it d} quarks, the number of the {\it d} quark
must be about two times of that of {\it u} quarks to ensure the
electrical neutrality. On the other hand, the finite size effect
including the contribution of the surface tension and the curvature,
which has been shown to be very important to the properties of
strangelets \cite{BJ87,Madsen934}, and of great interest in the
study of nucleus, neutron stars, and heavy ion collisions
\cite{Bodm71}, has not yet been taken into account for the isospin
asymmetric quark matter
\cite{Nishida04,SS01,KT013,BCPR03,KTV03,FBO03,LV034}. It is thus
imperative to investigate the variation behavior of the isospin
asymmetry and the chiral symmetry breaking in finite size quark
matter (or quark droplet). Since the Nanbu--Jona-Lasinio (NJL) model
\cite{NJL61} at quark level \cite{VW91,Buballa05} involves the
information of chiral symmetry breaking and restoration, and of
which the interaction among quarks is easy to deal with, in this
paper we take the NJL model and the multiple reflection expansion
(MRE) approximation \cite{BJ87,Madsen934,KH03,YHT05,BB70} to study
the finite size effect on the isospin asymmetric quark matter. For
simplicity, we take only the {\it u}, {\it d} quark matter into
account.

In the two-flavor NJL model, the Lagrangian is written as
\begin{equation}
\label{eq1} {\mathscr{L}} = \overline q (i\gamma ^\mu \partial _\mu
- m_0 + \mu \gamma _0 )q + {\mathscr{L}}_{int} ,
\end{equation}

\noindent where $m_0 $ is the current quark mass. For the flavors
{\it u} and {\it d}, one usually takes approximation $m_u = m_d =
m_0 $. The interaction part is
\begin{equation}
\label{eq2} {\mathscr{L}}_{int} = G[(\overline q q)^2 + (\overline
q i\gamma _5 \overrightarrow \tau q)^2] \, .
\end{equation}

In order to describe the isospin symmetry breaking, we introduce
different chemical potential for {\it u}, {\it d} quark,
respectively,
\begin{equation}
\label{eq3}
\mu = \left( {{\begin{array}{*{20}c}
 {\mu _u } \hfill & 0 \hfill \\
 0 \hfill & {\mu _d } \hfill \\
\end{array} }} \right)\, .
\end{equation}
It is obvious that this breaks the isospin symmetry, and the
breaking can be manifested by the isospin chemical potential $\mu _I
= \mu _u - \mu _d $, so different quark condensates $\phi _u =
\left\langle {\overline u u} \right\rangle $ and $\phi _d = \langle
{\overline d d} \rangle $ should be considered. The quark condensate
is generally given by
\begin{equation}
\label{eq4} \langle\overline{q}q\rangle=-i\int\frac{d^4
p}{(2\pi)^4}trS(p) \, ,\\
\end{equation}
where $S(p)$ is the dressed quark propagator and the trace is on the
Dirac and color spaces.
 After some derivations, the thermodynamic potential at finite
temperature $T$ and chemical potential $\mu_{f} \, (f = u, d)$ can
be written as
\begin{equation}
\label{eq5} \Omega (T,\mu _u , \mu _d ,\phi _u ,\phi _d ) = 2G(\phi
_u ^2 + \phi _d ^2) + \sum\limits_{f = u,d} {\Omega _{M_f } (T,\mu
_f )}   \, ,
\end{equation}
\noindent with
$$\displaylines{\hspace*{2cm} \Omega _{M_f } (T,\mu _f ) = - 2N_c \int
\frac{d^3k}{(2\pi )^3}\left\{ E_{k,f} + T\ln [1 + \exp ( -
\frac{1}{T}(E_{k,f} - \mu _f ))] \right. \hfill{} \cr
\hspace*{6cm}
\left. + T\ln [1 + \exp (-\frac{1}{T}(E_{k,f} + \mu _f ))] \right\}
\, , \hfill{} \cr }$$           %  (6)

\noindent where $E_{k,f} = \sqrt {M_f ^2 + k^2} $, $M_f $ is the
constituent quark mass and can be given at the mean field level as
\begin{equation}
\label{eq6}   M_f = m_0 - 2G\phi _f    \, .
\end{equation}

For the {\it u}, {\it d} quark system in a spherical bubble with
radius $R$ (or referred as a quark droplet), the chemical potential
is dependent on $R$, so we can write the Lagrangian as
\begin{equation}
\label{eq7} {\mathscr{L}} = \overline q (i\gamma ^\mu \partial _\mu
- m_0 + \mu (R) \gamma _0 )q + {\mathscr{L}}_{int} ,
\end{equation}
where $\mu (R)=\mu\rho $, the concrete form of $\rho$ will be given
later. Because the droplet (or bubble) has a finite surface and the
curvature provides a pressure difference between the inner and the
outer parts of the droplet, not only the quark matter in the
droplet, but also the surface and the curvature contribute to the
thermodynamical potential of the system \cite{OM93}. To incorporate
all these effects, an approach denoted as multiple reflection
expansion approximation has been developed
\cite{BJ87,Madsen934,KH03,YHT05,BB70}. In the multiple reflection
expansion approximation, for a droplet composing of quarks with
flavor $i$, the thermodynamical quantities can be derived from a
density of states in the form
\begin{equation}
\label{eq8}
\frac{dN_i}{dk}=6\Big[\frac{k^2V}{2\pi^2}+f_S\Big(\frac{k}{m_i}\Big)kS
+f_C\Big(\frac{k}{m_i}\Big)C+...\Big] \, ,
\end{equation}
where $V$ is the volume of the droplet, $S=4\pi R^2$ and $C=8\pi R$
are the area and the extrinsic curvature of the surface of the
droplet. $f_{S} \left( {\frac{k}{m}} \right)$, $f_{C} \left(
{\frac{k}{m}} \right)$ are the contributions to the density of the
states from the surface, the curvature of the droplet, respectively,
and can be given explicitly as \cite{BJ87,Madsen934}
\begin{equation}
\label{eq9} f_{S} \!\Big( {\frac{k}{m}} \Big) = - \frac{1}{8\pi }
\left( {1 - \frac{2}{\pi }\arctan \Big( {\frac{k}{m}} \Big)}
\right),
\end{equation}

\begin{equation}
\label{eq10} f_{C}\! \Big( {\frac{k}{m}} \Big) = \frac{1}{12\pi
^2}\left[ {1 - \frac{3k}{2m}\left( {\frac{\pi }{2} - \arctan \Big(
{\frac{k}{m}} \Big)} \right)} \right].
\end{equation}
Then the density of states of the $i$ flavor quarks can be given in
the multiple reflection expansion approximation as
\begin{equation}
\label{eq11}\rho= \rho _{MRE} (k,m,R) = 1 + \frac{6\pi ^2}{kR}
f_{S}\!\Big( {\frac{k}{m}} \Big) + \frac{12\pi ^2}{(kR)^2}
f_{C}\!\Big( {\frac{k}{m}} \Big)\, .
\end{equation}
From this expression we know that, if we do not consider the
contributions from the surface and the curvature, the Lagrangian in
Eq.~(\ref{eq7}) has the general form used in studying infinite quark
matter in the NJL model.

By using the multiple reflection expansion approximation, we obtain
then the density of states as $(k^2\rho _{MRE} ) / 2\pi ^2$ and the
thermodynamic potential can be given as

$$\displaylines{\hspace*{5mm} \Omega (T,\mu _u, \mu _d, \phi _u , \phi _d )
= \sum\limits_{f =u,d} { \frac{(M_f - m_0 )^2}{2G} } - 2N_c
\sum\limits_{f = u,d} \int \frac{\rho_{MRE} k^2dk}{2\pi ^2} \Big\{
E_{k,f} \hfill{} \cr \hspace*{42mm} + T\ln [1 + \exp ( -
\frac{1}{T}(E_{k,f} - \mu _f ))] + T\ln [1 + \exp ( -
\frac{1}{T}(E_{k,f} + \mu _f ))] \Big\} \, .  \hfill{(12)} \cr }
$$
\noindent{In the zero-temperature limit $T  \to 0 $}
$$\displaylines{\hspace*{1mm} \Omega (T,\mu _u ,\mu _d ,\phi _u ,\phi _d )
= \sum\limits_{f = u, d} {\frac{(M_f \!- \! m_0 )^2}{2G}} - 2N_c
\int_{k_F ^u}^\Lambda { \frac{E_{k,u} \rho _{MRE}  k^2dk}{2\pi ^2} }
- 2N_c \mu _{u} \int_0^{k_{F}^{u}} { \frac{\rho _{MRE} k^2 dk}{2\pi
^2}} \hfill{} \cr \hspace*{50mm}  - 2N_c \int_{k_F ^d}^\Lambda {
\frac{ E_{k,d} \rho _{MRE}k^2 dk}{2\pi ^2} - 2N_c \mu _{d}
\int_0^{k_F ^d} { \frac{\rho _{MRE} k^2 dk}{2\pi ^2}} } \, ,
\hfill{(13)} \cr } $$

\noindent where $\mu _u = \sqrt {(M_u )^2 + (k_F ^u)^2} $, $\mu _d =
\sqrt {(M_d )^2 + (k_F ^d)^2} $; $k_F ^u$, $k_F ^d$ are the Fermi
momentums of {\it u}, {\it d} quark, respectively.

Consequently we can obtain the {\it u}, {\it d} quark numbers
\setcounter{equation}{13}
\begin{equation}
\label{eq14} N_{u} = \mbox{ }V n_u = 2N_c V \int_0^{k_F ^u}
{\frac{\rho _{MRE} k^2dk}{2\pi ^2}}    \, ,
\end{equation}

\begin{equation}
\label{eq15} N_{d} = V n_d = 2N_c V \int_0^{k_F ^d} {\frac{\rho
_{MRE} k^2dk}{2\pi ^2}} \, ,
\end{equation}

\noindent where $V = 4\pi R^3 / 3$ is the volume of the droplet.
$n_f = - \frac{\partial \Omega }{\partial \mu _f }$ is the quark
number density. For given $N_u$ and $N_d$, the Fermi momentum of the
{\it u}, {\it d} quark can be determined, respectively.

If the system (droplet) is stable, the thermodynamic potential must
satisfy the stationary conditions
\begin{equation}
\label{eq16} \frac{\partial \Omega }{\partial M_u } = 0,  \qquad
\qquad  \frac{\partial \Omega }{\partial M_d } = 0 \, .
\end{equation}

After some derivation we obtain the constituent quark mass
\begin{equation}
\label{eq17} M_u = m_0 + 2GN_c \partial _m \int_{k_F ^u}^\Lambda
{\rho _{MRE} \frac{k^2E_{k,u} dk}{2\pi ^2} + 2GN_c \mu _u \partial
_m \int_0^{k_F ^u} {\rho _{MRE} \frac{k^2 dk}{2\pi ^2}} } \, ,
\end{equation}

\begin{equation}
\label{eq18} M_d = m_0 + 2GN_c \partial _m \int_{k_F ^d}^\Lambda
{\rho _{MRE} \frac{k^2E_{k,d} dk}{2\pi ^2} + 2GN_c \mu _d \partial
_m \int_0^{k_F ^d} {\rho _{MRE} \frac{k^2 dk}{2\pi ^2}} } \, ,
\end{equation}

\noindent Then we can fix the chemical potentials
\begin{equation} \label{eq19} \mu _u = \sqrt {(M_u )^2 + (k_F ^u)^2} ,
\qquad \mu _d = \sqrt {(M_d )^2 + (k_F ^d)^2}  \, ,
\end{equation}

\noindent and the isospin chemical potential
\begin{equation} \label{eq20}
\mu _I = \mu _u - \mu _d \, .
\end{equation}

By varying the radius $R$ of the droplet, we can investigate the
finite size effect on the isospin symmetry breaking and chiral
symmetry restoration.

As a numerical example we take a charge neutral droplet with total
quark number $9000$, and the ratio of {\it u}, {\it d} quark number
1:2. For the current quark mass, we take $m_0 =6$~MeV. For the
cutoff and the coupling constant, we take $\Lambda =590$~ MeV and
$G\Lambda ^2 =4.7$, with which the pion decay constant and pion mass
($f_{\pi} = 93$~MeV, $m_{\pi} =140$~MeV) are reproduced well
\cite{YHT05}.

\begin{figure}[htbp]
\begin{center}
\includegraphics[scale=1]{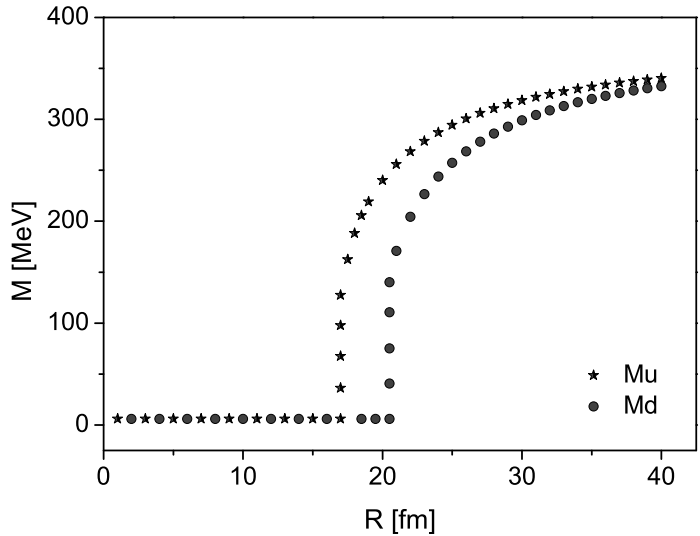}
\caption{\label{fig:cond-r} The dynamical quark masses of $u$, $d$
quarks as a function of the droplet radius $R$. }
\end{center}
\end{figure}

Since the dynamical quark mass can usually be taken to identify the
chiral symmetry breaking and isospin asymmetry, we investigate the
variation of the masses of {\it u} and {\it d} quarks at first. By
solving Eqs.~(\ref{eq14})-(\ref{eq18}), we obtain the dynamical
quark mass $M_f \; (f=u, d)$ as a function of radius $R$ of the
droplet. The result is illustrated in Fig.~1. From Fig.~1 we can
notice that the constituent quark mass $M_u $ is larger than $M_d $
at the same radius if the radius of the droplet is neither very
small nor extremely large. It means that the isospin symmetry is
broken. Whereas with the increase of $R$, the difference between
$M_u $ and $M_d $ becomes small. It indicates that, when the radius
$R$ of the droplet is large, the quark number density decreases
rapidly, and hence the isospin asymmetry becomes small. Meanwhile,
as the radius decreases (i.e., the quark number density increases
since the quark number is fixed), the dynamical mass of the quarks
can descend to current quark mass suddenly. It shows that the chiral
symmetry is restored except for the explicit breaking. The figure
also shows that the critical radius for the chiral symmetry of {\it
u} quark to be restored ($R=17$~fm) is smaller than that of {\it d}
quark ($R=20.5$~fm). When R is smaller than $17$ fm, the dynamical
masses of both {\it u} quark and {\it d} quark descend to current
quark masses. It manifests that the isospin symmetry can be
restored. In order to understand the phase transition in a more
clear way, we also study the quark condensates as a function of
baryon number density. The numerical result is illustrated in
Fig.~2. Fig.~2 shows evidently that the critical density of the
chiral symmetry restoration of {\it u} quark is larger than that of
{\it d} quark. More concretely, the critical density of {\it d}
quark is smaller than the normal nuclear matter density, whereas,
that of {\it u} quark is near the normal nuclear matter density. All
the results show that the contributions of the surface and the
curvature play important roles for the droplet with a small radius.

\begin{figure}[htbp]
\begin{center}
\includegraphics[scale=1]{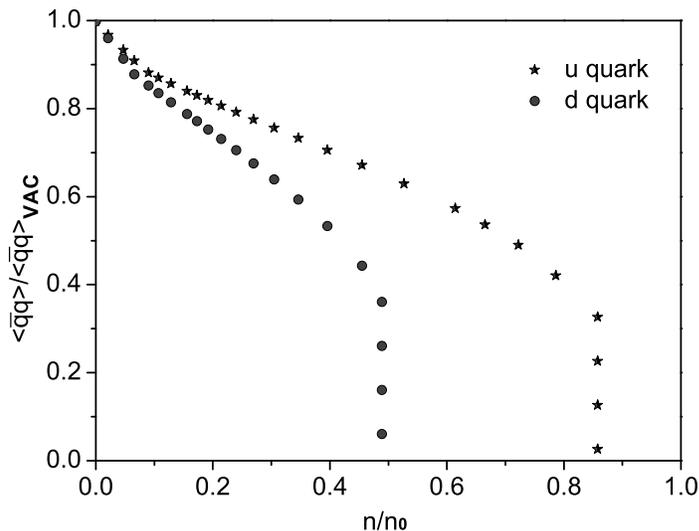}
\caption{\label{fig:condensate-b} Quark condensate as a function of
baryon number density. ${\langle\overline{q}q\rangle}_{VAC} $ refers
to the quark condensate with $R\rightarrow\infty$.}
\end{center}
\end{figure}

\begin{figure}[htbp]
\begin{center}
\includegraphics[scale=1]{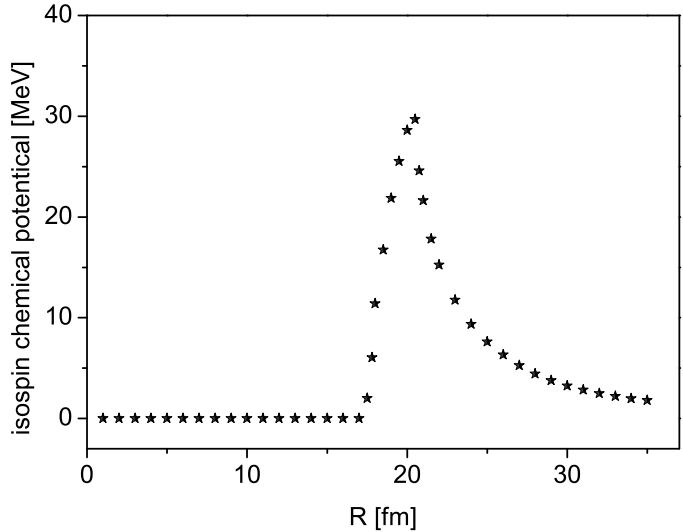}
\caption{\label{fig:iscp-r} The isospin chemical potential as a
function of the droplet radius $R$.}
\end{center}
\end{figure}

Then, we discuss the isospin chemical potential $\mu _I $ in the
droplet. By solving Eqs.~(14)-(20), we obtain the function $\mu _I $
versus the droplet radius. The result is shown in Fig.~3. It clearly
shows that when $R<17$~fm, the isospin chemical potential $\mu _I$
has the minimal value $0$, and it increases to the maximal value
when $R=20.5$~fm, then decreases with the increase of $R$. When the
radius $R$ is larger than $500$~fm, the difference between $\mu _u $
and $\mu _d $ is less than 1 percentage of the current quark mass.
This means that the isospin symmetry can be restored when the quark
number density is small. The zero isospin chemical potential at very
small radius shows that the isospin symmetry can also be restored if
the quark number density is quite large. Another thing important we
recognize is that, if the droplet takes a size so that only the
chiral symmetry of {\it u} quark could be restored, the isospin
chemical potential takes the maximal value about $29.7$~MeV in the
present example of the droplet. It is obvious that such a value is
much smaller than the critical isospin chemical potential for pion
condensate ($\mu _I ^c=m_\pi $ \cite{BCPR03,HZ05}) to occur. It
indicates that it may be difficult for pion condensate to happen in
the droplet at zero temperature.

The quark droplets may possibly be formed in the QCD phase
transition in the early universe or in astro-objects, for example,
the main ingredient in the inner part of pulsars is quark matter
composed of {\it u}, {\it d} quarks due to the high pressure
\cite{OM94}. Furthermore, some works once proposed that the new
phase of quark matter may emerge as droplets in nuclear matter at
low density \cite{HPS93}, even appears as {\it u-d} quark stars
(P-stars) \cite{Cea034,Cea05}. Our present results indicate that, as
the density of the {\it u-d} quark matter in the droplet is not very
small, the chiral symmetry can be restored, {\it i.e.}, the chiral
phase transition takes place. Since the phase transition influences
the equation of state of the matter drastically, the pressure, the
moment of inertia and other characteristics of the matter may change
suddenly. As shown in Figs.~1 and 2, when the density of the matter
in our present example is about $0.85n_{0}$ (the radius of the
droplet is about $17$~fm), the dynamical mass of {\it u} quark
changes from $127$~MeV to $6$~MeV. Such a sudden change may induce a
quake and a glitch for the astro-objects. And the light emission
strength may also be changed abruptly. When the density $n \in
(0.48n_0, 0.85n_0)$ (corresponding to the present example with
radius $17$~fm$ < R < 20.5$~fm), only the dynamical mass of {\it u}
quark changes. With a further decrease of the baryon number density,
the dynamical mass of {\it d} quark also changes suddenly. Then
another glitch may appear in pulsars. These phenomena may be taken
to identify the effect of chiral phase transition in the droplet of
{\it u} and {\it d} quarks \cite{GR05,Cea05}. The brightest giant
flare from soft gamma-ray repeaters and anomalous X-ray pulsars and
other observations may also be taken as the observable evidences to
signal the chiral phase transition effect. It is certain that
further studies are necessary to carry out a practical analysis. On
the other hand, the Figs.~1 and 2 manifest that, in a quite large
region of nuclear matter density, the isospin symmetry is broken,
and hence the dynamical mass of {\it u} quark is larger than that of
{\it d} quark. In the view point of bag models and soliton models of
a hadron, the effective mass of a proton (composing of 2 {\it u}
quarks and 1 {\it d} quark) in nuclear matter $m_{p}^{*}$ is larger
than that of a neutron (consisting of 1 {\it u} quark and 2 {\it d}
quarks) $m_{n}^{*}$. Together with the result given in QCD sum rules
\cite{Drukarev04}, we can recognize that the result $m_{p}^{*} >
m_{n}^{*}$ given in relativistic approaches
\cite{Liu02,Li041,Faes05} has solid microscopic foundation. Our
present result of the isospin asymmetry provides then a clue to
solve the controversial problem of the nucleon effective mass
splitting in nuclear matter
\cite{Liu02,Li041,Faes05,Li042,Ma04,Rizzo045} on the theoretical
side. It is now rather important since experiments have not yet
given any conclusion \cite{Rizzo045}.

In summary, by taking the two-flavor NJL model and the multiple
reflection expansion approximation, we have studied the phase
transition of a finite size droplet with {\it u} and {\it d} quarks.
We find that, the isospin symmetry is preserved and the chiral
symmetry is broken if the radius of the droplet is extremely large
(or the matter density is very small). With the radius changes from
infinite to about several hundred fms, both the chiral symmetry and
the isospin symmetry are broken. In such a case, the dynamical quark
mass $M_u $ is larger than $M_d $ at the same radius $R$. It
provides a clue that the effective mass of proton in nuclear matter
may be larger than that of neutron. If the radius of the droplet is
small enough (less than $20.5$~fm for $N=9000$), the chiral symmetry
can be restored. And the critical radius for {\it u}-quark is
smaller than that for {\it d}-quark. Meanwhile, possible
observations to identify the chiral phase transition are proposed in
some astronomical phenomena. In addition, the maximal isospin
chemical potential in the case of zero temperature is not large
enough so as to be comparable with the critical isospin chemical
potential for pion condensate to emerge. However in our present
study, we have not taken the flavor mixing \cite{Buballa05} into
account, where the dynamical quark mass $M_f$ ($f= u, d$) depends
not only on the condensate $\phi_{f}$ but also on $\phi _{f'}$.
Meanwhile the finite temperature effect has not yet been included
either. Moreover the experimental or astronomical observables
identifying the phase transition need also detailed investigations.
The related studies are under progress.

\bigskip

This work was supported by the National Natural Science Foundation
of China under contract Nos. 10425521 and 10135030, the Major State
Basic Research Development Program under contract No. G2000077400
and the Research Fund for the Doctoral Programme of Higher Education
of China under grant No 20040001010. One of the authors (Y.X. Liu)
would acknowledge the support of the Foundation for University Key
Teacher by the Ministry of Education, China.

%\newpage

\bigskip
\bigskip
\bigskip

\end{document}